\documentclass[prl,amsmath,amssymb,dvips,twocolumn,showpacs,preprintnumbers]{revtex4}
\usepackage{epsfig,bbold,algorithm,algorithmic,bm}
\usepackage{bm}

\newcommand{\beq}{\begin{equation}}
\newcommand{\eeq}{\end{equation}}
\newcommand{\bea}{\begin{eqnarray}}
\newcommand{\eea}{\end{eqnarray}}
\newcommand{\Tr}{\text{Tr}}
\newcommand{\kcut}{k_{\text{max}}^{}}                     
\newcommand{\pictsize}{0.875}
\newcommand{\etal}{{\it et al.,\;}}

\begin{document}

\preprint{
LA-UR-11-12070,
INT-PUB-11-058,
NT@UW-11-30
}

\title{Equation of state of the unitary Fermi gas: An update on lattice calculations}
\author{Joaqu\'{\i}n E. Drut$^1$, Timo A. L\"ahde$^2$\footnote{Present address: Institut f\"ur Kernphysik (Theorie),
Forsch\-ungszentrum J\"ulich,
D-52425 J\"ulich, Germany}, Gabriel Wlaz\l owski$^3$, and Piotr Magierski$^{3,4}$}
\affiliation{$^1$Theoretical Division, Los Alamos National Laboratory, Los Alamos, NM 87545--0001, USA}
\affiliation{$^2$Helsinki Institute of Physics, P.O. Box 64, FI-00014 University of Helsinki, Finland}
\affiliation{$^3$Faculty of Physics, Warsaw University of Technology, ulica Koszykowa 75, 00-662 Warsaw, Poland}
\affiliation{$^4$Department of Physics, University of Washington, Seattle, Washington 98195-1560, USA}

\date {\today}

\begin{abstract}
The thermodynamic properties of the unitary Fermi gas (UFG) have recently been measured to unprecedented
accuracy at the MIT. In particular, these measurements provide an improved understanding of the regime
below $T/\epsilon_F^{} \simeq 0.20$, where a transition into a superfluid phase occurs. In light of this development, we present
an overview of state-of-the-art auxiliary field quantum Monte Carlo (AFQMC) results for the UFG at finite
temperature and compare them with the MIT data for the energy, chemical potential, and density. These AFQMC
results have been obtained using methods based on the hybrid Monte Carlo (HMC) algorithm, which was first
introduced within the context of lattice QCD.
\end{abstract}


\maketitle


The unitary Fermi gas~(UFG) is defined as a two-component many-fermion system
in the limit of short interaction range $r_0^{}$ and large $s$-wave scattering length~$a$, such that
$0 \leftarrow k^{}_F r^{}_0 \ll 1 \ll k^{}_F a \rightarrow \infty$, with $k^{}_F \equiv (3 \pi^2 n)^{1/3}_{}$ being the Fermi momentum 
and $n$ being the particle number density (we choose units such that $\hbar = k_{B}^{} = 1$). 
The UFG also saturates the unitarity bound on the quantum-mechanical scattering 
cross section $\sigma_0^{}(k) \leq 4\pi / k^2$, where $k$ is the relative momentum of the colliding particles. 
The UFG features special properties that arise from the fact that it is characterized by a single 
length scale, given by the interparticle distance $\sim k_F^{-1}$, independently of the details of the interaction. 
While the thermodynamic
properties of the UFG are {\it universal}~\cite{Universality}, the lack of a readily accessible dimensionless expansion parameter 
renders the UFG a challenging many-body problem. Since the proposal of the UFG as a model for dilute neutron
matter by Bertsch~\cite{GFBertsch} and its realization in ultra-cold atom experiments 
(see Ref.~\cite{FirstExperiments} for a review of the experimental situation), the UFG has received
widespread attention across multiple disciplines, ranging from
atomic physics~\cite{Atoms} to the study of nuclear matter~\cite{Nuclei} and
relativistic heavy-ion collisions~\cite{RHIC}.

On the experimental side, the presence of a superfluid phase in the UFG below $T/\epsilon_F^{} \simeq 0.15$ was 
demonstrated directly a few years ago through the creation of an Abrikosov vortex lattice under 
rotation~\cite{Vortex}. However, a direct 
thermodynamic signature of the transition was not unambiguously established until the recent high-precision 
measurement at MIT of the
equation of state~(EoS) of a homogeneous two-component UFG
over a large temperature range~\cite{MIT_exp}. These measurements were performed on trapped $^6$Li atoms 
(using a Feshbach resonance to tune the system to the unitary limit), which
enabled a detailed study of the compressibility, density, and pressure of the UFG.
In addition, greatly refined empirical results were obtained for the associated critical temperature 
$T_c^{}/\epsilon_F^{} = 0.167(15)$ as well as for the ``Bertsch parameter'' $\xi = 0.376(5)$, which
characterizes the ground state of the UFG. As precision data are now available for the energy, chemical potential,
and density of the UFG in a wide temperature range, an opportunity presents itself to compare
these measurements with calculations in various theoretical frameworks. Here, we focus on comparing 
the MIT data with the most recent auxiliary field quantum Monte Carlo (AFQMC) results.

The Hamiltonian that captures the physics of the unitary limit can be written on a spatial lattice as 
\beq
\hat H \equiv \!\!\!
\sum_{\bm{k},\lambda=\uparrow,\downarrow} \!\!
\frac{k^2}{2m} \,
\hat a^\dagger_\lambda(\bm{k}) \, \hat a_\lambda^{}(\bm{k})
-\,g \sum_i \hat n_\uparrow^{}(\bm{r}_i^{}) \, \hat n_\downarrow^{}(\bm{r}_i^{}),
\eeq
where $\lambda$ denotes the spin projection, $m$ is the fermion mass (we also set $m = 1$), 
and $g$ is the coupling constant.
The creation and annihilation operators satisfy fermionic anticommutation relations, and 
$\hat{n}_\lambda^{}(\bm{r}_i^{}) \equiv\hat{a}_\lambda^\dagger(\bm{r}_i^{})\, \hat{a}_\lambda^{}(\bm{r}_i^{})$ 
denotes the number density operator at lattice position $\bm{r}_i^{}$ for spin projection~$\lambda$.
The thermodynamic equilibrium properties are obtained from the partition function
\beq\label{Z_original}
\mathcal Z \equiv \Tr\,\exp[-\beta(\hat H \!-\! \mu \hat N)],
\eeq
where $\hat{N}$ is the total particle number operator, 
$\mu$ is the chemical potential, and $\beta \equiv 1/T$ is the inverse temperature.

To evaluate expectation values of observables numerically, we followed
the path-integral approach presented extensively in Ref.~\cite{BDM}, with recent improvements described 
in Ref.~\cite{DL_contact}. The system is placed on a cubic spatial lattice of extent $L=N_x^{} l$
with periodic boundary conditions. The lattice spacing $l$ (henceforth set to unity) and extent $L$ provide
natural ultraviolet~(UV) and infrared~(IR) momentum cutoffs, given by $\kcut=\pi/l$ and $k_0^{}=2\pi/L$, 
respectively. 
The imaginary-time evolution operator $\exp[-\beta(\hat{H}-\mu \hat{N})]$ is expanded using a Trotter-Suzuki 
decomposition with temporal
lattice spacing $\tau$, and the interaction is represented by means of a Hubbard-Stratonovich (HS) 
transformation~\cite{HST}. 
As we focus on the spin-symmetric case, the fermion sign problem is absent. The resulting path integral formulation is an 
exact representation of Eq.~(\ref{Z_original}) up to finite-volume and discretization effects, which 
may be controlled by varying the spatial lattice volume $V \equiv N_x^3$ and density $n$. The thermodynamic 
and continuum limits are recovered as $V \to \infty$ and $n \to 0$, respectively. The latter requires great care, as too low 
densities may introduce shell effects.
As our lattice formulation is very similar to that of
Ref.~\cite{BDM}, referred to here as determinantal Monte Carlo (DMC),
we shall restrict ourselves to describing three modifications
which significantly improve the results.

First, the bare-lattice coupling constant $g$ corresponding to the 
unitary regime is determined by means of L\"uscher's formula~\cite{Luescher} as in Ref.~\cite{LeeSchaefer}. 
This procedure yields $g \simeq 5.14$ in 
the unitary limit. Our lattice Hamiltonian contains $g$ as the sole parameter characterizing the interaction. 
Finite-range effects are induced by the presence of the UV cutoff of the lattice. In order to minimize such 
discretization effects, the dilute limit should be approached as closely as possible.
Recent theoretical developments~\cite{Endresetal,EKLN,JED} have explored the use of improved transfer matrices and operators, 
with multiple parameters tuned to unitarity. The implementation of such methods is an objective of future AFQMC calculations.

Second, we use a compact, continuous HS transformation
referred to as ``Type~4'' in Ref.~\cite{DeanLee}, which found it superior with respect to acceptance rate, decorrelation
and signal-to-noise properties than the more conventional unbounded and discrete HS transformations~\cite{Hirsch}. 

Third, we update the HS auxiliary field $\sigma$ using hybrid
Monte Carlo (HMC), which combines molecular dynamics
(MD) evolution of $\sigma$ with a Metropolis accept-reject step~\cite{DuaneGottlieb}.
The result is the determinantal hybrid Monte Carlo (DHMC)
algorithm, introduced in Ref.~\cite{DL_contact}. In DHMC, global MD updates of $\sigma$ take place via introduction of a momentum field 
$\pi$ conjugate to $\sigma$, such that the dynamics is given by the Hamiltonian
\beq
\label{H_DHMC}
\mathcal H \equiv \sum_{i} \frac{\pi_i^2}{2} 
-\ln \det \left[ \left (\mathbb{1} + \mathcal U_{}^{} [\sigma] \right)^2\right],
\eeq
where $\mathcal U [\sigma]$ encodes the dynamics of the fermion degrees of freedom (for more details on 
${\mathcal U}$, see {\it e.g.} Ref.~\cite{BDM}).
The DHMC algorithm produces greatly enhanced decorrelation between successive MC samples for all temperatures and lattice sizes, 
and removes the necessity to spend an increasing
number of decorrelation steps at larger $V$ (note that the computational cost of a full sweep of the lattice scales as 
$\sim V$ in a local algorithm such as DMC). This is replaced by a fixed number of operations, typically of $\mathcal{O}(10^2)$, 
required to produce one MD ``trajectory'', independently of $V$. 

\begin{figure}[t]
\begin{center}
\includegraphics[width=\pictsize\columnwidth]{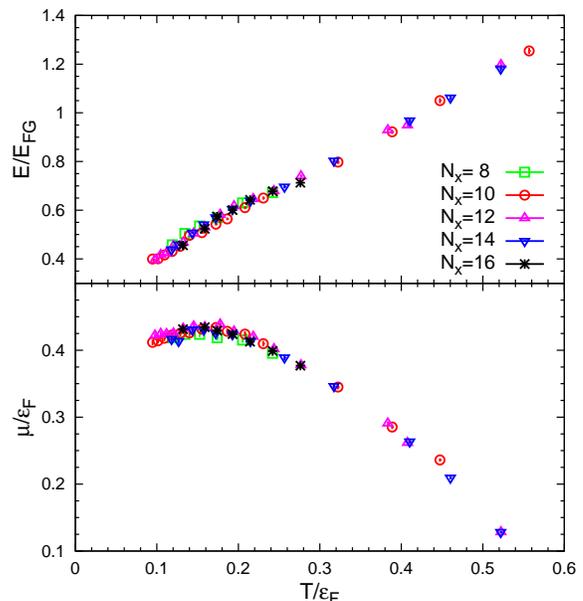}
\end{center}
\vspace{-.8cm}
\caption{(Color online) AFQMC results for (top) the total energy
$E$ in units of the energy of a free Fermi gas and (bottom) the chemical
potential $\mu$ in units of the Fermi energy as a function of $T/\epsilon_F^{}$.
Results are shown for $N_x^{} = 8$ (green squares), $N_x^{} = 10$ (red circles), $N_x^{} = 12$ (purple triangles),
$N_x^{} = 14$ (blue inverted triangles), and $N_x^{} = 16$ (black asterisks), where $V \equiv N_x^3$ is the (spatial) 
lattice volume.
}
\label{Fig:dhmc_results}
\end{figure}
\begin{figure}[t]
\begin{center}
\includegraphics[width=\pictsize\columnwidth]{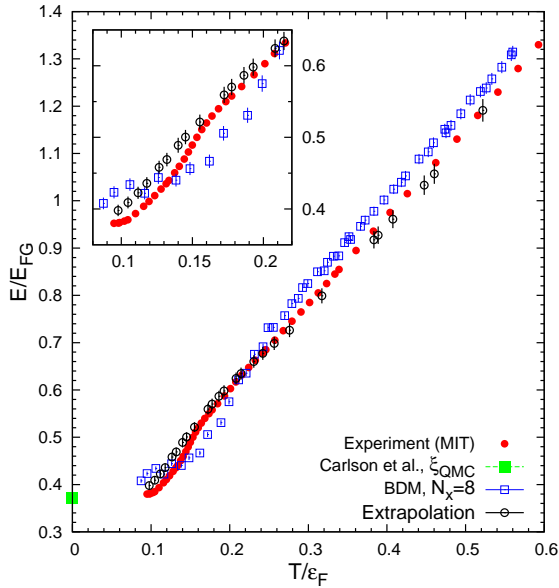}
\end{center}
\vspace{-.8cm}
\caption{(Color online) Energy $E/E_\text{FG}^{}$ (red dots), as obtained by Ku {\it et al.}~\cite{MIT_exp}. Our AFQMC results
extrapolated to infinite volume are shown by open black circles. The results for $N_x^{} = 8$ (open blue squares) were obtained 
with the DMC algorithm in Ref.~\cite{BDM}. The green square shows the QMC result of Ref.~\cite{Carlsonetal} for $\xi$ at $T = 0$. 
The inset
shows the vicinity of the superfluid phase transition at $T_c^{}/\epsilon_F^{} \simeq 0.15$.  
}
\label{Fig:comp1}
\end{figure}
\begin{figure}[t]
\begin{center}
\includegraphics[width=\pictsize\columnwidth]{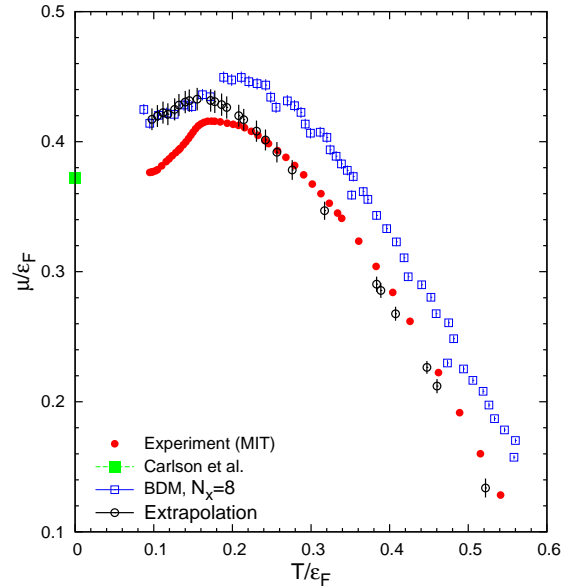}
\end{center}
\vspace{-.8cm}
\caption{(Color online) Chemical potential $\mu$ in units of $\epsilon_F^{}$ 
as measured by Ku {\it et al.}~\cite{MIT_exp}. The notation
for the AFQMC results is identical to Fig.~\ref{Fig:comp1}.
The solid green square shows the result of Ref.~\cite{Carlsonetal} assuming $\mu/\epsilon_F^{} (T = 0) = \xi$.
}
\label{Fig:comp2}
\end{figure}

The DHMC algorithm allows for an extension of the AFQMC analysis beyond the 
capabilities of DMC, which is currently limited to $N_x^{} \simeq 10$ in
the spatial lattice extent, $n \simeq 0.1$ in particle number density, and $N \simeq 100$ in the particle number.
The improved scaling of the CPU time has allowed us to study lattices up to $N_x^{} = 16$, while
simultaneously maintaining a relatively large number of particles, $N \simeq 45,75,110$, and $160$ for $N_x^{} = 10,12,14$ and 
$16$, respectively, which corresponds to densities in the range $n \simeq 0.040-0.045$. The $N_x^{} = 8$ data 
(which are not identical to Ref.~\cite{BDM}), correspond to $N \simeq 35$ and $n \simeq 0.070$, which
was not reduced further in order to avoid shell effects. We generated $\simeq 200$ uncorrelated snapshots of $\sigma$ 
for each value of $T/\epsilon_F^{}$, which yields a statistical uncertainly of $\simeq 1\%$ for the 
observables. Expressions for the AFQMC computation of the latter were obtained by differentiating ${\mathcal Z}$
with respect to $\beta$ and $\mu$ as conventional in thermodynamics, with the exception that ${\mathcal Z}$
is replaced by its discretized form in terms of the HS field.

In Fig.~\ref{Fig:dhmc_results}, we present AFQMC results (for various lattice sizes) for the total energy $E$ in units of the energy of a 
free Fermi gas $E_\text{FG}^{} = 3/5 N \epsilon_F^{}$, and for the chemical potential $\mu$
in units of $\epsilon_F^{}=k_{F}^{2}/2m$ as a function 
of $T/\epsilon_F^{}$. Before comparison with the MIT data~\cite{MIT_exp}, we performed an extrapolation to the 
infinite volume limit $N_{x}\rightarrow\infty$. This required interpolation of the data series for each value of $N_x^{}$, as the physical
temperature $T/\epsilon_F^{}$ is not known beforehand. Apart from this minor complication, the extrapolation to infinite volume is
greatly facilitated by the lack of a systematic variation in the results with the lattice volume above $N_x^{} \simeq 10$.
Since we have not performed
an extrapolation to the continuum limit (which requires $n\rightarrow 0$), our results may be affected to some degree by systematic
errors due to the effective range $r_{\text{eff}}^{}$. Our results currently reach $k_F^{} r_{\text{eff}}^{} \simeq 0.3$ which is
non-negligible and may produce significant deviations, in particular at low $T/\epsilon_F^{}$ as shown by Carlson {\it et al.}~\cite{Carlsonetal}.

\begin{figure}[t]
\begin{center}
\includegraphics[width=\pictsize\columnwidth]{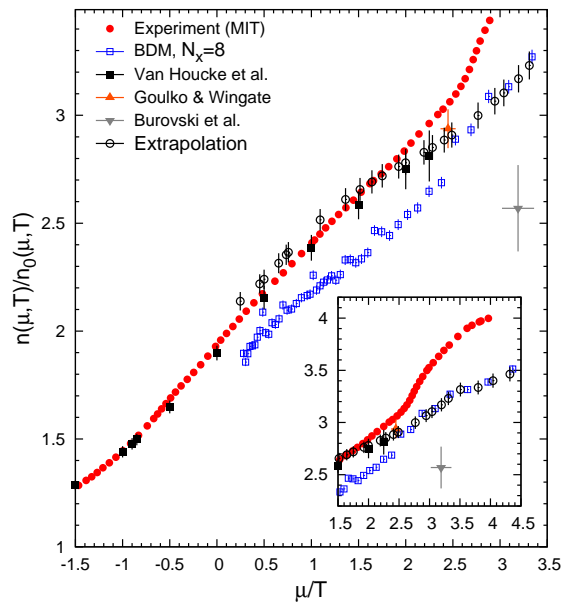}
\end{center}
\vspace{-.5cm}
\caption{(Color online) 
Density $n(\mu,T)$ of the UFG (red circles) as obtained by Ku {\it et al.}~\cite{MIT_exp}, 
normalized to the density $n_0^{}(\mu,T)$ of a non-interacting Fermi gas. The notation for 
the AFQMC results is identical to Fig.~\ref{Fig:comp1}.
The diagrammatic MC results of Refs.~\cite{Burovski,Goulko} (solid up and down triangles) and the Bold Diagrammatic MC results
of Ref.~\cite{VanHoucke} are shown as well (solid squares). The inset shows the
vicinity of the superfluid phase transition at $T_c^{}/\epsilon_F^{} \simeq 0.15$.  
}
\label{Fig:comp3}
\end{figure}

In Fig.~\ref{Fig:comp1}, we compare our AFQMC results (extrapolated to infinite volume) with the 
measured energy $E$ of the homogeneous UFG. The overall agreement is
satisfactory throughout the range of temperatures studied. At low $T/\epsilon_F^{}$, AFQMC slightly overpredicts
the experimental data. Our new results show a noticeable improvement over the results of Ref.~\cite{BDM} with 
$N_x^{} = 8$, likely due to decreased finite-volume and effective-range effects.
In contrast to 
the case of $E/E^{}_\text{FG}$, our results for $\mu/\epsilon_F^{}$ in Fig.~\ref{Fig:comp2} deviate noticeably from 
experiment at low $T/\epsilon_F^{}$, where $\mu/\epsilon_F^{} \simeq 0.38$ at $T/\epsilon_F^{} \simeq 0.1$. 
AFQMC overpredicts this by $\simeq 5\%$, which clearly exceeds the statistical uncertainty. 
However, the larger lattices used here represent a dramatic improvement over Ref.~\cite{BDM},
in particular above $T_c^{}/\epsilon_F^{} \simeq 0.15$. Nevertheless, the discrepancy below $T_c^{}/\epsilon_F^{}$ cannot be
accounted for at present. In Fig.~\ref{Fig:comp3}, we show the particle number density relative to 
the temperature-dependent density of the noninteracting Fermi gas. Again, a discrepancy at low $T/\epsilon_F^{}$
is found, which is analogous to that observed for $\mu/\epsilon_F^{}$.

While the agreement between our AFQMC calculation and the data of Ref.~\cite{MIT_exp} is satisfactory in general,
notable discrepancies persist. We have achieved a significant reduction 
of the density from $n \simeq 0.1$ to $n \simeq 0.04$, with a concomitant decrease in discretization (finite-range) effects.
Nevertheless, since finite-range effects scale as $\sim n^{1/3}$, this still only implies an effective reduction from $n^{1/3} \simeq 0.46$ 
to $\simeq 0.34$. The possibility that the discrepancies between our AFQMC data and experiment are due to residual finite-range
effects can therefore not be ruled out at present.

As the region where the discrepancies are largest appears to be at very low $T/\epsilon_F^{}$ (at least for $E/E_\text{FG}^{}$ 
and $\mu/\epsilon_F^{}$), the task of performing calculations at significantly lower values of $n^{1/3}_{}$ for such temperatures
is extremely demanding, indeed largely beyond the capabilities of extant algorithms. In this situation, accounting for the finite-range
effects by improving the transfer matrix (as in Refs.~\cite{Endresetal,EKLN}) provides 
a systematic way to remove the finite-range effects from both the action and the observables at a given density, 
without modifying the temperature scale of the calculation. Preliminary results have appeared in 
Ref.~\cite{JED}. Another source of error under investigation is the Trotter-Suzuki step $\tau$. This was found to be a small effect
for the Tan contact in Ref.~\cite{DL_contact}, as well as for the energy in Ref.~\cite{EKLN}. 

In spite of these 
shortcomings, the introduction of HMC into the AFQMC study of the UFG has largely solved the issue 
of sufficiently large spatial lattice dimension $N_x^{}$ and particle number $N$, which in turn 
has allowed calculations with a large particle number at lower densities. DHMC studies for 
$N_x^{} > 16$ are in progress. These improvements will also apply to calculations away from the 
unitary limit. Finally, we would like to stress that the AFQMC method is entirely {\it ab initio}: once 
the coupling $g$ is fixed by solving the two-body problem, no tuning with respect to experiment is required.
While a more sophisticated analysis of the systematic errors cannot be provided at this point in time, the fact that theory and 
experiment agree reasonably well for both $E/E_\text{FG}^{}$ and $\mu/\epsilon_F^{}$ over a wide
range of temperatures is both remarkable and encouraging.


We thank M.~Zwierlein for making the MIT data available, 
and F.~Werner for providing us with the BDMC results.
We are also indebted to A.~Bulgac for instructive discussions and a careful reading of the manuscript.
We acknowledge support under U.S. DOE Grant No. DE-FC02-07ER41457, and
Contract No. N N202 128439 of the Polish Ministry of Science. This study was supported, in part, by the Vilho, Yrj\"o, and Kalle 
V\"ais\"al\"a Foundation of the Finnish Academy of Science and Letters and by the Magnus Ehrnrooth, the
Waldemar von Frenckell, and the Ruth and Nils-Erik Stenb\"ack foundations of the Finnish Society of Sciences and Letters.  One of the authors (G.W.) acknowledges the Polish Ministry of Science for
the support within the program ``Mobility Plus - I edition'' under Contract No. 628/MOB/2011/0.
Part of the computer time for this study was provided by 
the Interdisciplinary Centre for Mathematical and Computational Modeling (ICM) at the University of Warsaw. 


\end{document}